\documentclass[journal,10pt,twocolumn]{IEEEtran}
\usepackage{amssymb,flushend,relsize,graphicx, placeins,amsmath,psfig,cite,float,setspace}
\usepackage{xcolor}
\usepackage{tabularx}
\usepackage{hyperref}
\usepackage{subfigure}
\usepackage{algorithm}
\usepackage{algpseudocode}

\newcommand{\bi}{\begin{itemize}}
\newcommand{\ei}{\end{itemize}}

\newcommand{\be}{\begin{enumerate}}
\newcommand{\ee}{\end{enumerate}}
\newcommand{\bd}{\begin{description}}
\newcommand{\ed}{\end{description}}
\newcommand{\bc}{\begin{center}}
\newcommand{\ec}{\end{center}}
\newcommand{\bt}{\begin{tabbing}}
\newcommand{\et}{\end{tabbing}}
\newcommand{\bfig}{\begin{figure}}
\newcommand{\efig}{\end{figure}}
\newcommand{\beq}{\begin{equation}}
\newcommand{\beqarr}{\begin{eqnarray}}
\newcommand{\beqarrn}{\begin{eqnarray*}}
\newcommand{\eeq}{\end{equation}}
\newcommand{\eeqarr}{\end{eqnarray}}
\newcommand{\eeqarrn}{\end{eqnarray*}}
\newcommand{\bflr}{\begin{flushright}\vspace{-0.2in}}
\newcommand{\eflr}{\end{flushright}}
\newcommand{\bsub}{\begin{subequations}}
\newcommand{\esub}{\end{subequations}}
\newcommand{\barr}{\begin{array}}
\newcommand{\earr}{\end{array}}
\newcommand{\nn}{\nonumber}

\def\undb#1{\mbox{\bf{#1}}}

\def\dn{\stackrel{\scriptscriptstyle \triangle}{=}}

\begin{document}
\bstctlcite{bstctl:nodash}

\title{\huge{Optimal Multi-Level ASK Modulations for RIS-Assisted Communications with Energy-Based Noncoherent Reception}}
\author{Sambit Mishra, \IEEEmembership{Student Member, IEEE}, Soumya P. Dash, \IEEEmembership{Senior Member, IEEE},\\ and George C. Alexandropoulos, \IEEEmembership{Senior Member, IEEE}
\thanks{S. Mishra and S. P. Dash are with the School of Electrical Sciences, Indian Institute of Technology Bhubaneswar, Argul, Khordha, India (email: 21ec01021@iitbbs.ac.in, soumyapdashiitbbs@gmail.com).}
\thanks{G. C. Alexandropoulos is with the Department of Informatics and Telecommunications, National and Kapodistrian University of Athens, Panepistimiopolis Ilissia, 15784 Athens, Greece, and also with the Department of Electrical and Computer Engineering, University of Illinois Chicago, IL 60601, USA (email: alexandg@di.uoa.gr).}

}
{}
\maketitle
\begin{abstract}
This paper investigates the performance of one- and two-sided amplitude shift keying (ASK) modulations in noncoherent single-input single-output (SISO) wireless communication systems assisted by a reconfigurable intelligent surface (RIS). Novel noncoherent receiver structures are proposed based on the energy of the received symbol and the choice of the modulation scheme for data transmission. The system's performance is assessed in terms of the symbol error rate (SER) and an optimization framework is proposed to determine the most effective one- and two-sided ASKs to minimize the SER, while adhering to average a transmit power constraint. Two scenarios based on the availability of the statistical characteristics of the wireless channel are explored: a) the transceiver pair has complete knowledge of the channel statistics, and b) both end nodes possess knowledge of the statistics of the channel gain up to its fourth moment, and novel algorithms are developed to obtain optimal ASKs for both of them. Extensive numerical evaluations are presented showcasing that there exists a threshold signal-to-noise ratio (SNR) above which the optimal ASKs outperform the traditional equispaced ASKs. The dependencies of the SER performance and the SNR threshold on various system parameters are assessed, providing design guidelines for RIS-assisted noncoherent wireless communication systems with multi-level ASK modulations.
\end{abstract}
\IEEEpeerreviewmaketitle
\section{Introduction}
Recent advancements in wireless communication systems are paving the way for sixth-generation (6G) wireless networks, which are anticipated to support highly demanding use cases, such as holographic communications, extended reality, tactile internet, advanced on-board communications, integrated sensing and communications, as well as global ubiquitous connectivity \cite{JiHa21, HMIMO, DT_RIS, DaJoSa22}. These visionary applications are expected to drive a tremendous surge in data traffic and complexity, requiring substantial energy and hardware investments \cite{WaYo23,DISAC}. To address these challenges, various physical-layer technologies have been explored, including extremely massive multiple-input multiple-output (MIMO)~\cite{XYA2024}, millimeter-wave and terahertz (THz) communications, and ultra-dense network architectures, some even within the framework of fifth-generation (5G) networks. However, network energy consumption and hardware footprint constitute significant challenges that still remain to be efficiently addressed.

A promising technology for 6G systems that offers potential solutions to blockage issues and coverage enhancement in a cost-effective and energy-efficient manner is the reconfigurable intelligent surface (RIS)~\cite{RIS_multi,Ja24,RIS_tutorial,RIS_Marconi,DaKa24,DaJoAi22, BhAiDa23, VaDaAc23}. An RIS consists of miniature metamaterials of programmable electromagnetic responses that can intelligently manipulate electromagnetic waves to enhance signal strength at the receiver and optimize communication links. By adjusting the phases of impinging signals, RISs create adaptive wireless environments that can redirect these signals to desired directions/points in space~\cite{RIS_deployment,RIS_pervasive}. This capability is typically optimized to mitigate path loss, interference, and shadowing issues across frequencies ranging from low bands to THz~\cite{RIS_THz}, enhance network coverage~\cite{MeZh23}, improve device localization —even in scenarios without access points—, and provide energy-efficient physical-layer security \cite{RIS_localization,RIS_secrecy}. Recently, RISs have been integrated with unmanned aerial vehicles to boost energy and spectrum efficiencies \cite{GeDo20, MaMu21, MoAiICC20, MoAiTWC23}, utilized in wireless power transfer and energy harvesting across different frequency bands \cite{MEH, MoAiPIMRC22}, employed in index modulation systems \cite{BaDaGh24, BaDaKa24, BaDaKaMa24, BaDaJoGh23}, and in free-space optical communication systems \cite{NaSc21}.

The performance analysis of RIS-assisted wireless systems has garnered significant attention in the research community over recent years. The authors in \cite{VeMa22} investigated the outage probability and packet error rate of a dual-hop free-space optical radio frequency communication system aided by an RIS under phase errors. In \cite{YeGu20}, a joint reflecting and precoding scheme was proposed for RIS-assisted MIMO communications to reduce the symbol error rate (SER). An RIS-based system to enhance the coverage of hybrid visible light communications in indoor environments was introduced in \cite{PaDaAi24}. To address scenarios where the direct link between a base station and a user is blocked, an RIS-assisted non-orthogonal multiple access scheme was proposed in \cite{SuJi22}, demonstrating that transmit power decreases linearly with the number of base station antennas and quadratically with the number of RIS elements. The performance of RIS-empowered systems under Nakagami-$m$ fading was analyzed in \cite{RIS_Nak} for both random and coherent phase configuration schemes. A deep reinforcement learning method for jointly optimizing RIS phase configurations and transmit precoding in a multi-user setup, considering quality-of-service constraints, was studied in \cite{YaXi21}. The impact of correlated Nakagami-$m$ channels on RIS-based communication was examined in \cite{AjDaRa23, AjDaRa22}, revealing that system performance improves with increasing RIS elements but degrades with higher channel correlation. Closed-form upper bounds and location-dependent expressions for the ergodic capacity over cascaded Rician fading channels were derived in \cite{RIS_rotation}. In \cite{WuZh19}, a joint active and passive beamforming algorithm utilizing instantaneous channel state information (CSI) was proposed for RIS-aided communication, showing that the power scaling relative to the number of RIS elements follows a squared law.

Although several such studies on the performance and optimization of RIS-assisted wireless systems have been carried out, their vast majority assume perfect CSI availability at the receiver. Typically, this becomes a challenging task owing to the increased overhead and complexity in channel estimation due to the passive and reflective nature of RISs \cite{RIS_tutorial,JAS2022}. To overcome the reliance on pilot-assisted channel estimation methods, noncoherent communication schemes have been studied, which eliminate the need for pilot signal transmission and simplify receiver design by removing the necessity for channel or phase estimation and compensation.
The benefits of noncoherent communications have recently inspired the design of noncoherent schemes for RIS-assisted wireless systems \cite{Se22, Kun_1, Kun_2, CaHuBa23, CaHuAn23, InWa24}. In \cite{Se22}, it was shown that the performance of RIS-assisted noncoherent MIMO communications improves with increasing numbers of RIS elements. A noncoherent demodulation scheme for differential modulation, combined with codebook-based beam training, for zero-overhead training was presented in \cite{Kun_1}. An RIS-assisted orthogonal frequency division multiplexing system using differential phase shift keying with random RIS configurations was analyzed in \cite{Kun_2}, showcasing superior performance over coherent demodulation in various mobility and spatial correlation scenarios. In \cite{CaHuBa23}, a noncoherent RIS-aided system using joint index keying and $M$-ary differential chaos shift keying was shown to achieve error performance comparable to coherent RIS-based spatial modulation systems. Significant error performance improvements for a code-index-modulation-aided differential chaos shift keying system incorporating an RIS were reported in \cite{CaHuAn23}. Very recently, \cite{InWa24} introduced a noncoherent RIS system with differential modulation, requiring low training complexity and exhibiting robustness against channel variations.

Although the afore-described studies on RIS-assisted noncoherent communications avoid the channel estimation step, the considered receiver structures still impose high energy and hardware requirements. Thus, the design of hardware-efficient reception for noncoherent RIS-assisted wireless systems remains an open challenge. Motivated by this research gap, this paper analyzes the performance of RIS-aided noncoherent single-input single-output (SISO) communication systems employing one- and two-sided amplitude shift-keying (ASK). We consider two scenarios: one where the statistics of the fading channels are completely available at the transceiver pair, and another where only the first four moments of the channel are known. The key contributions of the paper are summarized as:
\begin{enumerate}
\item Novel noncoherent energy-based reception structures for one- and two-sided ASK modulation are presented.
\item We formulate and solve a novel optimization framework for the system's SER minimization under constraints on the average transmit energy.
\item Novel algorithms to obtain SER-minimizing one- and two-sided ASK constellations for both considered scenarios of statistical channel availability are proposed.
\item Extensive numerical results are presented to demonstrate the efficacy of the proposed algorithms, showcasing our system's superior SER performance with the designed optimal ASKs at high signal-to-noise ratios (SNRs), as compared to traditional equispaced ASK signaling.
\end{enumerate}

The remainder of the paper is organized as follows. Section II describes the system model and introduces the proposed receiver structures for ASK modulations, while Section III presents our mathematical formulations for designing SER-minimizing constellations. Section IV includes the algorithms to solve the optimization problems for both cases of statistical channel availability. The paper's numerical evaluations validating the SER analysis and illustrating the structures of the optimal modulation schemes are included in Section V. Finally, Section VI concludes the paper.

\textit{Notations:} $[M]$ denotes the set ${1, 2, \dots, M}$ with $M$ being an integer. $\mathbb{C}^{n\times m}$ represents the set of all complex-valued matrices of size $n \times m$. For a matrix $\mathbf{H} \in \mathbb{C}^{n\times m}$, $H_{i,j}$ denotes the element in the $i^{th}$ row and $j^{th}$ column. For a vector $\mathbf{h} \in \mathbb{C}^{n\times 1}$, $h_i$ denotes the $i^{th}$ element. $\mathcal{CN}(\boldsymbol{\mu}, \mathbf{R})$ represents a circularly symmetric complex Gaussian random vector with mean vector $\boldsymbol{\mu}$ and covariance matrix $\mathbf{R}$. $\mathcal{N}(\mu, \sigma^2)$ denotes a real Gaussian random variable with mean $\mu$ and variance $\sigma^2$. $\mathbf{X}^H$ indicates the Hermitian transpose (conjugate transpose) of matrix $\mathbf{X}$. Finally, $\mathbf{1}_N$ is the $N \times 1$ vector of ones and $\mathbf{I}_N$ denotes the $N \times N$ $(N \geq 2)$ identity matrix.
\section{System Model}
Consider an RIS-assisted wireless communication system comprising of a single-antenna transmitter, a single-antenna receiver, and an RIS with $N$ elements of tunable reflection states. The transmitter employs a multi-level ASK modulation for data transmission, while the receiver performs symbol-by-symbol detection for noncoherent reception. Let $\undb{h}_1 \in \mathbb{C}^{N \times 1}$ denote the gain vector for the channel between the transmitter and the RIS and $\undb{h}_2 \in \mathbb{C}^{N \times 1}$ represent the gain vector of the wireless channel between the RIS and the receiver. We assume that the line-of-sight (LoS) link between the transmitter and receiver is blocked, making the RIS-assisted path the sole one establishing the communication link. Thus, the baseband received signal can be mathematically expressed as follows:
\begin{equation}
y = \undb{h}_{2}^{H} \mathbf{\Phi} \undb{h}_1 x + n,
\label{eq1}
\end{equation}
where $\mathbf{\Phi} \in \mathbb{C}^{N \times N}$ is the diagonal matrix consisting of the phase shifts of the reflecting elements of the RIS, $x \in \mathbb{R}$ is the transmitted ASK-modulated symbol, and $n \in \mathbb{C}$ is the additive white Gaussian noise (AWGN) with $n \sim \mathcal{CN}(0, 2 \sigma_n^2)$. Since the RIS creates a virtual LoS communication link when a direct link is unavailable, we consider that the channel gains between the transmitter-RIS and RIS-receiver links follow complex Gaussian distributions with non-zero means, implying that $\undb{h}_1 \sim \mathcal{CN} \left( \mu_{1} \mathbf{1}_N, \sigma_{h}^2 \mathbf{I}_N \right)$ and $\undb{h}_2 \sim \mathcal{CN}\left( \mu_{2}\mathbf{1}_N, \sigma_{h}^2\mathbf{I}_N \right)$.

The transmitter is considered to be a part of a base station, implying that it has sufficient hardware capability to estimate the uplink channel. This facilitates the RIS to employ the optimal phase shifts at its reflecting elements, implying that $(\mathbf{\Phi})_{n,n} = e^{\jmath(\angle h_{1,n} - \angle h_{2,n})}$ for $n \in [N]$ \cite{RIS_tutorial}. Furthermore, considering that the transmitted signal is real and that the optimal phase shifts result in a real-valued composite channel gain, we focus on the real part of the received signal for our analysis, which is obtained as
\begin{equation}
\Re \left\{ y \right\} \triangleq y_r = \left( \sum_{n=1}^N \left| h_{1,n} \right| \left| h_{2,n} \right| \right) x + n_r,
\label{eq2}
\end{equation}
where $n_r = \Re \left\{n \right\} \sim {\mathcal{N}} \left(0, \sigma_n^2 \right)$. Assuming a large $N$ and applying the central limit theorem, the sum $\sum_{n=1}^N \left| h_{1,n} \right| \left| h_{2,n} \right|$ can be approximated as a Gaussian random variable, i.e.:
\begin{equation}
\sum_{n=1}^N \left| h_{1,n} \right| \left| h_{2,n} \right| \sim \mathcal{N}\left( \mu_{f}, \sigma_{f}^2 \right),
\label{eq3}
\end{equation}
where the moments are given by
\begin{subequations}
\begin{align}
\mu_{f} &= \alpha \sigma_{h}^2, \quad \sigma_{f}^2 = \beta \sigma_{h}^4,
\label{eq4a}
\end{align}
with the parameters $\alpha$ and $\beta$ defined as follows:
\begin{align}
\alpha &\dn \frac{N \pi}{4} L_{1/2} \left( - K_1 \right) L_{1/2} \left( - K_2 \right),  \\
\beta &\dn N \left[ \left(1+K_1 \right) \left(1+K_2 \right) - \frac{\pi^2}{16} L_{1/2}^2 \left(-K_1 \right) L_{1/2}^2 \left(-K_2 \right) \right],\notag
\label{eq4b}
\end{align}
\end{subequations}
with $K_1 \dn \left| \mu_1 \right|^2/\sigma_h^2$ and $K_2 \dn \left| \mu_2 \right|^2/\sigma_h^2$ being the Rician factors of the transmitter-RIS and RIS-receiver channels, respectively, and $L_{1/2}(\cdot)$ denoting the Laguerre polynomial of order $1/2$~\cite{GrRy07}. In the special case of Rayleigh fading channels (i.e., $\mu_1 = \mu_2 = 0$), the expressions for $\alpha$ and $\beta$ simplify to:
\begin{equation}
\alpha \big|_{\mu_1=\mu_2=0} = \frac{N \pi}{4}, \quad \beta \big|_{\mu_1=\mu_2=0} = \frac{N \left( 16 - \pi^2 \right)}{16}.
\label{eq5}
\end{equation}
\subsection{One- and Two-Sides ASK Signaling}
The transmitter maps information onto a multi-level ASK constellation. In this paper, we consider the following two types of $M$-level ASK modulation:
\begin{enumerate}
\item \textit{One-sided ASK}:
\begin{equation}
x_m = \sqrt{E_m}, \quad m \in [M],
\label{eq6a}
\end{equation}
\item \textit{Two-sided ASK}:
\begin{equation}
x_m = \begin{cases}
 -\sqrt{E_{\frac{M}{2} - m +1}}, & \text{if } 1 \leq m \leq \frac{M}{2} \\
 \sqrt{E_{m - \frac{M}{2}}}, & \text{if } \frac{M}{2} < m \leq M
\end{cases},
\label{eq6b}
\end{equation}
\end{enumerate}
where $E_m$ (for $m \in [M]$) represents the energy of the $m$-th symbol in the ASK constellation. We assume that all symbols are transmitted with equal probability. Without loss of generality, we arrange the one-sided ASK constellation such that the amplitude levels are in increasing order, i.e., $\sqrt{E_m} < \sqrt{E_{m+1}}$ for all $m \in [M-1]$. Similarly, for the two-sided ASK modulation, we have $\sqrt{E_m} < \sqrt{E_{m+1}}$ for $m \in \left[\frac{M}{2}-1\right]$. To this end, the average energy of the multi-level ASK constellation is calculated as follows:
\begin{equation}
E_{\text{av}} = \frac{1}{M} \sum_{m=1}^M x_m^2,
\label{eq7}
\end{equation}
and we define the energy codebook used by the transmitter as:
\begin{equation}
\mathcal{E} \dn \begin{cases}
\left\{ E_m: m \in [M] \right\}, & \text{for one-sided ASK} \\
\left\{ E_m: m \in \left[\frac{M}{2}\right] \right\}, & \text{for two-sided ASK}
\end{cases}.
\label{eq9}
\end{equation}
Owing to the statistics of the channel and noise, the SNR of each $m$-th symbol ($m = 1,\ldots,M$) is defined as: 
\beq
\Gamma_m \triangleq \frac{(2\beta + \alpha^2) E_m \sigma_h^4}{\sigma_n^2},
\label{eq11}
\eeq
yielding the following expression for the average SNR of the system:
\beq
\Gamma_{\rm av} = \frac{(2\beta + \alpha^2)E_{\rm av}\sigma_h^4}{\sigma_n^2},
\label{eq12}
\eeq
where using \eqref{eq7} and \eqref{eq9} it holds that $E_{\rm av} = \frac{1}{M} \sum_{m=1}^M E_m$.
\subsection{Reception for One-Sided ASK}
Consider that the transmitter sends a one-sided ASK modulated symbol $x_m$ with power $p_m$ for $m \in [M]$, as defined in \eqref{eq6a}. To design an energy- and hardware-efficient receiver, we deploy an energy detector. To this end, upon obtaining $y_r$, the receiver computes the following quantity:
\beqarr
&& \! \! \! \! \! \! \! \! \! \! \! \! \! \! \! \!
\! \! \! \! \! \! \! \!
\frac{|y_r|^2}{\mathbb{E} \left[\left( \sum\limits_{n=1}^{N} |h_{1,n}| |h_{2,n}| \right)^2 \right]}
= \frac{|y_r|^2}{\sigma_h^4 \left( \alpha^2 + \beta \right)} \nn \\
&& \qquad \qquad \quad
= \frac{\left| \left( \sum\limits_{n=1}^{N} |h_{1,n}| |h_{2,n}| \right) x + n_r \right|^2}
{\sigma_h^4 \left( \alpha^2 + \beta \right)}.
\label{eq13}
\eeqarr
Then, using the channel statistics, it creates the decision regions as $\mathcal{D} = \left\{ I_m\right \}_{m=1}^{M}$, where each $I_m$ corresponds to a power level $p_m$, and in turn, to a signal in the set of the one-sided ASK constellation. In the sequel, the receiver estimates the transmitted symbol index, $m$, as:
\begin{equation}
\hat{m} \in \left\{ m: \frac{|y_r|^2}{\sigma_h^4 \left( \alpha^2 + \beta \right)} \in I_{m} \right\} \, .
\label{eq14}
\end{equation}
Hence, the probability of error when the $m$-th symbol with power $p_m$ is transmitted, $P_e(p_m)$, and the average symbol error rate (SER), $P_s$, for a given constellation size $M$ are defined as:
\begin{equation}
P_e \left(p_m \right) \triangleq \Pr \left( \hat{m} \neq m \right),
\quad P_s \triangleq \frac{1}{M} \sum_{m=1}^{M} P_e \left(p_m \right).
\label{eq15}
\end{equation}
\subsection{Reception for Two-Sided ASK}
Similar to the case of one-sided ASK, when the transmitter uses two-sided ASK modulation, the receiver first computes the expression in (\ref{eq13}) upon collecting $y_r$. Since two-sided ASK consists of both positive and negative symbols, it also computes the following quantity:
\begin{equation}
\hat{y}_r = \frac{y_r}{|y_r|}.
\label{eq16}
\end{equation}
It is noted that, for two-sided ASK, the power level $p_m$ is allocated to two symbols in the constellation diagram. 

Following the latter computations, the receiver first decodes the transmitted symbol index as in (\ref{eq14}) to link it to a symbol with power level $p_m$ where $m \in \left[\frac{M}{2}\right]$. Then, it determines the transmitted symbol index $m \in [M]$ using the following rule:
\begin{equation}
\hat{m} = \begin{cases}
m, & \text{if } \hat{y}_r < 0, \\
 m + \frac{M}{2}, & \text{if } \hat{y}_r > 0
\end{cases}.
\label{eq17}
\end{equation}
\section{Proposed Optimization Framework}
The SER of the considered RIS-assisted noncoherent system can be obtained through the decision boundaries presented previously in Sections~II.B and~II.C for the one- and the two-sided ASK constellations, respectively. In this section, we focus on designing the characteristics of both ASK constellations with the goal of minimizing the SER performance under practical constraints in the total/average energy available at the transmitter for symbol transmission. 
\subsection{Energy Constraints for One- and Two-Sided ASK}
For the traditional equispaced one-sided ASK, we have $E_1 = 0$ and $E_m = 4(m-1)^2$ for $m \geq 2$, resulting in the following expression for the total energy of the constellation:
\begin{equation}
\sum_{m=1}^{M} E_m = \frac{2 M (M-1)(2M-1)}{3}.
\label{eq18}
\end{equation}
To ensure that the designed optimal constellation does not exceed the average energy of the traditional one-sided ASK, we impose the following energy constraint:
\begin{equation}
\frac{1}{M} \sum_{m=1}^{M} E_m \leq \frac{2}{3}(M-1)(2M-1).
\label{eq19}
\end{equation}

For the conventional two-sided ASK constellation, where $E_1 = 4$ and $E_m = 4m^2$ for $m \geq 2$, we have the total energy of the constellation to be given as follows:
\begin{equation}
\sum_{m=1}^{M/2} E_m = \frac{(M+2)M(M+1)}{6} \, .
\label{eq20}
\end{equation}
To design an optimal constellation within the average energy limit of the traditional two-sided ASK, the following energy constraint need to be satisfied:
\begin{equation}
\frac{2}{M} \sum_{m=1}^{M/2} E_m \leq \frac{1}{3}(M+1)(M+2) \, .
\label{eq21}
\end{equation}
\subsection{ASK Design Problem Formulation}
Using the energy constraints in (\ref{eq19}) and (\ref{eq21}), we formulate the following optimization problem for both one- and two-sided ASK constellation:
\begin{align}\label{eq22}
\underset{\mathcal{E}, \mathcal{D}}{\text{min}}\,\log(P_s) \,\,
\text{s.t.} \,\, \frac{1}{\mathcal{M}}\!\sum_{m=1}^{\mathcal{M}} E_m \leq \mathcal{C}, \, E_m \geq 0 \, \forall m \in [\mathcal{M}],
\end{align}
where $\mathcal{E}$ is the energy codebook defined in (\ref{eq9}), $\mathcal{D}$ is the decision regions as given in Section~II.B, and:
\begin{equation}
\mathcal{M} = \begin{cases}
M, & \text{for one-sided ASK}  \\
\frac{M}{2}, & \text{for two-sided ASK} 
\end{cases},
\label{eq23}
\end{equation}
whereas, 
\begin{equation}
\mathcal{C} = \begin{cases}
\frac{2}{3}(M-1)(2M-1), & \text{for one-sided ASK} \\
\frac{1}{3}(M+1)(M+2), & \text{for two-sided ASK} 
\end{cases}.
\label{eq24}
\end{equation}
It is noted that the considered logarithm of the SER in (\ref{eq22}) yields a monotonically increasing function, which facilitates the solution of the optimization problem.

Assuming that the $m^{\text{th}}$ symbol $x_m$ with energy $E_m$ is transmitted, and using (\ref{eq11}), it is deduced that:
\beqarr
\mathbb{E}\left[\frac{|y_r|^2}{\sigma_h^4 \left( \alpha^2 + \beta \right)}\right]
\! \! \! \!
&=& \! \! \! \! \frac{\mathbb{E}\left[ \left|
\left(\sum\limits_{n=1}^{N} |h_{1,n}| |h_{2,n}| \right) \sqrt{E_m}
+ n_r \right|^2 \right]}
{\sigma_h^4 \left( \alpha^2 + \beta \right)} \nn \\
&=& \! \! \! \! \frac{E_m \mathbb{E}
\left[ \left(\sum\limits_{n=1}^{N} \left| h_{1,n} \right|
\left| h_{2,n} \right| \right)^2 \right]
+ \mathbb{E} \left[ n_r^2 \right]}
{\sigma_h^4 \left( \alpha^2 + \beta \right)} \nn \\
&=& \! \! \! \! E_m + \tilde{\sigma}_n^2,
\label{eq25}
\eeqarr
where we have used the definition:
\begin{equation}
\tilde{\sigma}_n^2 \triangleq \frac{\sigma_n^2}{\sigma_h^4 \left( \alpha^2 + \beta \right)} \, .
\label{eq26}
\end{equation}
To this end, we define the receiver's constellation points as:
\begin{equation}
r \left( E_m \right)
\triangleq \mathbb{E} \left[\frac{|y_r|^2}
{\sigma_h^4 \left( \alpha^2 + \beta \right)} \right]
= E_m + \tilde{\sigma}_n^2.
\label{eq27}
\end{equation}
Further, we define the decoding regions $I_m$ as:
\begin{equation}
I_m \triangleq \left(r \left( E_m \right) - d_{l,m},
r \left( E_m \right) + d_{r,m} \right),
\label{eq28}
\end{equation}
where $d_{l,m} > 0$ and $d_{r,m} > 0$ represent the maximum allowable deviations to the left and right of $r \left( E_m \right)$, respectively. Thus, an upper bound on the symbol error probability $P_s$ can be obtained as follows~\cite{ScalLaws}:
\begin{equation}
P_s \leq \frac{1}{\mathcal{M}}
\sum_{m=1}^{\mathcal{M}}
\left(e^{-I_{r,m}(d_{r,m})} + e^{-I_{l,m}(d_{l,m})}\right) ,
\end{equation}
where $I_{l,m}(d)$ and $I_{r,m}(d)$ denote the left and right rate functions, respectively, and are defined as:
\begin{subequations}
\begin{align}
I_{l,m}(d) &\triangleq \sup_{\theta \geq 0}
\left(\theta d - \log(M_m(-\theta)) \right),
\label{eq30a} \\
I_{r,m}(d) &\triangleq \sup_{\theta \geq 0}
\left(\theta d - \log(M_m(\theta)) \right),
\label{eq30b}
\end{align}
\end{subequations}
with $\sup_{x \in A} f(x)$ being the supremum of $f(x)$ over the set $A$, while $M_m(\theta)$ is the moment-generating function of $u_m$:
\begin{equation}
M_m(\theta) = \mathbb{E} \left[ e^{\theta u_m} \right]
\label{eq31}
\end{equation}
with the random variable $u_m$ defined as follows:
\begin{equation}
\begin{split}
u_m &\triangleq \frac{|y_r|^2}{\sigma_h^4 \left( \alpha^2 + \beta \right)}
- \mathbb{E} \left[ \frac{|y_r|^2}
{\sigma_h^4 \left( \alpha^2 + \beta \right)}\right] \\
&= \frac{\left| \left(\sum\limits_{n=1}^{N}
\left|h_{1,n} \right| |h_{2,n}| \right) \sqrt{E_m} + n_r\right|^2}
{\sigma_h^4 \left( \alpha^2 + \beta \right)}
- r \left( E_m \right).
\end{split}
\label{eq32}
\end{equation}

As shown in~\cite{ScalLaws}, the rate functions in (\ref{eq30a}) and (\ref{eq30b}) have the following properties:
\begin{enumerate}
    \item 
    \begin{equation}
    \lim_{d \rightarrow 0} \frac{I_{r,m}(d)}{d^2} = \lim_{d \rightarrow 0} \frac{I_{l,m}(d)}{d^2} = \frac{1}{2 \mathbb{E}[u_m^2]}.
    \label{eq33}
    \end{equation}
    \item They are non-negative, convex, and monotonically increasing functions of $d$ for $d > 0$ and fixed non-negative $E_m$. They are also monotonically decreasing functions of $E_m$ for fixed positive $d$.
    \item They satisfy $I_{l,m}(0) = I_{r,m}(0) = 0$ for any non-negative value of $p_m$.
\end{enumerate}
Hence, for any $d > 0$, the minimum of $I_{l,m}(d)$ and $I_{r,m}(d)$ represents the error exponent associated with the probability of incorrectly decoding $E_m$, when the receiver uses the interval $I_m$ as the decoding region. Thus, the error exponent for the SER of the considered system, denoted by $I_e$, is given as:
\begin{equation}
I_e \triangleq \min_{m \in [\mathcal{M}]}
\left(I_{l,m}(d_{l,m}), I_{r,m}(d_{r,m})\right).
\label{eq34}
\end{equation}
This implies that the optimization framework in (\ref{eq22}) for the design of the constellation points and the decoding regions can be reformulated in terms of the error exponent, as follows:
\begin{align}\label{eq35}
& \underset{\mathcal{P}, \mathcal{D}}{\text{max}}\, I_e \,\,\, \text{s.t.} 
\,\,\,\frac{1}{\mathcal{M}} \sum_{m=1}^{\mathcal{M}} E_m
\leq \mathcal{C},\, E_m \geq 0\, \forall m \in [\mathcal{M}],
\end{align}
with $\mathcal{M}$ and $\mathcal{C}$ given in (\ref{eq23}) and (\ref{eq24}), respectively. Note that this optimization formulation serves as a relaxation of the one in (\ref{eq22}), since its objective function provides an upper bound on that used in the initial formulation.
\section{SER-Optimal ASK Constellations}
To solve the optimization problem in (\ref{eq35}), we consider two distinct cases for the availability of statistical information about the channel at the communication ends. The first case assumes that the transceiver pair has perfect knowledge of the channel statistics, whereas, in the second case, only the first four moments of the fading distribution are known.
\subsection{Case 1: Perfect Knowledge of Channel Statistics}
Let us assume that the receiver has perfect knowledge of the statistics of the fading channel. With the exact channel distribution known, $M_m(\theta)$ can be determined at both the receiver and transmitter for any selected constellation point $E_m$. Therefore, the optimization problem in (\ref{eq35}) becomes:
\begin{align}\label{eq36}
    & \underset{\{E_m, d_{l,m}, d_{r,m}\}_{m \in [\mathcal{M}]}}{\text{max}} 
     \quad \min_{m \in [\mathcal{M}]}
     \left( I_{l,m}(d_{l,m}), I_{r,m}(d_{r,m}) \right)\nonumber \\
    & \text{s.t.} 
    \quad  \frac{1}{\mathcal{M}}\sum_{m=1}^{\mathcal{M}} E_m
    \leq \mathcal{C}, \, E_m \geq 0 \,
    \forall m \in [\mathcal{M}].
\end{align}
From the structures of the rate functions, i.e., the exponential function, the average energy of the transmitted symbols needs to be small to ensure a convergence when computing the solution of the optimization problem. Therefore, we use $\mathcal{C}$ given in (\ref{eq24}) and define the scaled-down energy $\tilde{E}_m$ and the scaled-down bounds $\tilde{d}_{l,m}$ and $\tilde{d}_{r,m}$ as follows:
\beq
\tilde{E}_m \triangleq \frac{E_m}{\mathcal{C}}, \quad
\tilde{d}_{l,m} \triangleq \frac{d_{l,m}}{\mathcal{C}}, \quad
\tilde{d}_{r,m} \triangleq \frac{d_{r,m}}{\mathcal{C}},
\label{eq37}
\eeq
which we replace to \eqref{eq36} yielding:
\begin{align}\label{eq38}
& \underset{\{\tilde{E}_m, \tilde{d}_{l,m},
\tilde{d}_{r,m}\}_{m \in [\mathcal{M}]}}
{\text{max}} 
\quad \min_{m \in [\mathcal{M}]}
\left( I_{l,m} \left( \tilde{d}_{l,m} \right),
I_{r,m} \left( \tilde{d}_{r,m} \right) \right)\nonumber \\
& \quad\text{s.t.}
\quad \frac{1}{\mathcal{M}}
\sum_{m=1}^{\mathcal{M}} \tilde{E}_m \leq 1,
\tilde{E}_m \geq 0 \,\forall m \in [\mathcal{M}].
\end{align}
\begin{algorithm}[!t]
\caption{One-Sided ASK Constellation for Case 1}\label{alg1}
\begin{algorithmic}[1]
\State $\left[\mathcal{E}, \mathcal{D} \right] \gets \Call{BisectionOneSided}{}$
\vspace{1mm}
\hrule
\vspace{1mm}
\Function{ConstellationDesign}{$t$}
    \State Initialize $\tilde{E}_1 \gets 0$, $\tilde{d}_{l,1} \gets 0$, $\tilde{d}_{r,M} \gets \infty$.
    \For{$m = 1$ to $M$}
        \If{$m = 1$}
            \State Set $\tilde{E}_1 \gets 0$.
        \Else
            \State Find smallest $\tilde{E}_m > \tilde{E}_{m-1} + \tilde{d}_{r,m-1}$
            \Statex \hspace{1.45cm} such that $I_{l,m} \left(\tilde{E}_m - \tilde{E}_{m-1} - \tilde{d}_{r,m-1} \right) = t$.
            \State Set $\tilde{d}_{l,m} \gets \tilde{E}_m - \tilde{E}_{m-1} - \tilde{d}_{r,m-1}$.
        \EndIf
        \If{$m \neq 1$}
            \State Find smallest $\tilde{d}_{r,m} > 0$ such that
            \Statex \hspace{1.45cm} $I_{r,m} \left(\tilde{d}_{r,m} \right) = t$.
        \EndIf
    \EndFor
    \State Assemble $\tilde{\mathcal{E}} \gets \left\{ \tilde{E}_m \right\}_{m=1}^{M}$.
    \State Assemble $\tilde{\mathcal{D}} \gets \left\{ 
           \left( \tilde{d}_{l,m}, \tilde{d}_{r,m} \right) \right\}_{m=1}^{M}$
    \State \Return $\tilde{\mathcal{E}}, \tilde{\mathcal{D}}$.
\EndFunction
\end{algorithmic}
\end{algorithm}
\begin{algorithm}[!t]
\caption{Two-Sided ASK Constellation for Case 1}\label{alg2}
\begin{algorithmic}[1]
\State $\left[ \mathcal{E}, \mathcal{D} \right] \gets \Call{BisectionTwoSided}{}$
\vspace{1mm}
\hrule
\vspace{1mm}
\Function{ConstellationDesign}{$t$}
    \State Initialize $\tilde{E}_0 \gets 0$, $\tilde{d}_{l,1} \gets 0$, $\tilde{d}_{r,M/2} \gets \infty$.
    \For{$m = 1$ to $M/2$}
        \State Find smallest $\tilde{E}_m > \tilde{E}_{m-1} + \tilde{d}_{r,m-1}$
        \Statex \hspace{0.93cm} such that
        $I_{l,m}\left(\tilde{E}_m - \tilde{E}_{m-1} - \tilde{d}_{r,m-1}\right) = t$.
        \State Set $\tilde{d}_{l,m} \gets \tilde{E}_m - \tilde{E}_{m-1} - \tilde{d}_{r,m-1}$.
        \If{$m \neq 1$}
            \State Find smallest $\tilde{d}_{r,m} > 0$ such that
            \Statex \hspace{1.45cm} $I_{r,m}\left(\tilde{d}_{r,m}\right) = t$.
        \EndIf
    \EndFor
    \State Assemble $\tilde{\mathcal{E}} \gets \left\{\tilde{E}_m\right\}_{m=1}^{M/2}$.
    \State Assemble $\tilde{\mathcal{D}} \gets \left\{ \left(\tilde{d}_{l,m}, \tilde{d}_{r,m}\right) \right\}_{m=1}^{M/2}$.
    \State \Return $\tilde{\mathcal{E}}, \tilde{\mathcal{D}}$
\EndFunction
\end{algorithmic}
\end{algorithm}
\begin{algorithm}[!t]
\caption{Bisection Method for One-Sided ASK}\label{3}
\begin{algorithmic}[1]
\Function{BisectionOneSided}{}
    \State Initialize $t_{\text{low}} \gets 0$, $t_{\text{high}} \gets \infty$.
    \State Set $\mathcal{C} \gets \dfrac{2}{3}\left(M - 1\right)\left(2M - 1\right)$.
    \Repeat
        \State Compute $t \gets \dfrac{t_{\text{low}} + t_{\text{high}}}{2}$.
        \State $\left[\tilde{\mathcal{E}}, \tilde{\mathcal{D}}\right] \gets \Call{ConstellationDesign}{t}$.
        \State Compute $S \gets \dfrac{1}{M} \sum_{m=1}^M \tilde{E}_m$.
        \If{$S < 1$}
            \State Update $t_{\text{low}} \gets t$.
        \Else
            \State Update $t_{\text{high}} \gets t$.
        \EndIf
    \Until{$|t_{\text{high}} - t_{\text{low}}| < \epsilon$ \textbf{and} $|S - 1| < \epsilon$}
    \State Set $\mathcal{E} \gets \left\{\mathcal{C} \tilde{E}_m \right\}_{m=1}^{M}$.
    \For{$m = 1$ to $M$}
        \State Set $I_m \gets \left(\mathcal{C} \tilde{E}_m + \sigma_n^2 - \mathcal{C} \tilde{d}_{l,m}, \right.$
        \Statex \hspace{2.5cm} $\left. \mathcal{C} \tilde{E}_m + \sigma_n^2 + \mathcal{C} \tilde{d}_{r,m} \right)$.
    \EndFor
    \State Collect $\mathcal{D} \gets \left\{ I_m \right\}_{m=1}^{M}$.
    \State \Return $\mathcal{E}, \mathcal{D}$
\EndFunction
\end{algorithmic}
\end{algorithm}
\begin{algorithm}[!t]
\caption{Bisection Method for Two-Sided ASK}\label{4}
\begin{algorithmic}[1]
\Function{BisectionTwoSided}{}
    \State Initialize $t_{\text{low}} \gets 0$, $t_{\text{high}} \gets \infty$.
    \State Set $\mathcal{C} \gets \dfrac{1}{3}\left(M + 1\right)\left(M + 2\right)$.
    \Repeat
        \State Compute $t \gets \dfrac{t_{\text{low}} + t_{\text{high}}}{2}$.
        \State $[\tilde{\mathcal{E}}, \tilde{\mathcal{D}}] \gets \Call{ConstellationDesign}{t}$.
        \State Compute $S \gets \dfrac{2}{M} \sum_{m=1}^{M/2} \tilde{E}_m$.
        \If{$S < 1$}
            \State Update $t_{\text{low}} \gets t$.
        \Else
            \State Update $t_{\text{high}} \gets t$.
        \EndIf
    \Until{$|t_{\text{high}} - t_{\text{low}}| < \epsilon$ \textbf{and} $|S - 1| < \epsilon$}
    \State Set $\mathcal{E} \gets \left\{\mathcal{C} \cdot \tilde{E}_m \right\}_{m=1}^{M/2}$.
    \For{$m = 1$ to $M/2$}
        \State Set $I_m \gets \left(\mathcal{C} \tilde{E}_m + \sigma_n^2 - \mathcal{C} \tilde{d}_{l,m}, \right.$
        \Statex \hspace{2.5cm} $\left. \mathcal{C} \tilde{E}_m + \sigma_n^2 + \mathcal{C} \tilde{d}_{r,m} \right)$.
    \EndFor
    \State Collect $\mathcal{D} \gets \left\{ I_m \right\}_{m=1}^{M/2}$.
    \State \Return $\mathcal{E}, \mathcal{D}$
\EndFunction
\end{algorithmic}
\end{algorithm}
The solution of the problem in~\eqref{eq38} is detailed in Algorithms~\ref{alg1}~and \ref{alg2} for one- and two-sided ASK constellations, respectively, and the constellation designs are based on the properties of the rate function.
To construct the optimal constellation, we first assume that a constellation with an error exponent $t$ exists that satisfies the scaled-down energy constraint. Starting with the minimum possible value for $\tilde{E}_1$, we choose the right-side boundary distance $\tilde{d}_{r,1}$ of the decoding region $I_1 = ( -\infty, r (\tilde{E}_1) + \tilde{d}_{r,1})$ such that the right rate function $I_{r,1} ( \tilde{d}_{r,1} )$ is at least $t$ on the right boundary. Following this, we choose the smallest value of $\tilde{E}_2$ such that $r(\tilde{E}_2) > r( \tilde{E}_1) + \tilde{d}_{r,1}$ and the left rate function $I_{l,2}(\tilde{d}_{l,2})$ is at least $t$ on the left boundary, where the left boundary distance is $\tilde{d}_{l,2} = r ( \tilde{E}_2) - r(\tilde{E}_1) - \tilde{d}_{r,1}$. Next, we find the right-side boundary distance $\tilde{d}_{r,2}$ of $I_2 = ( r ( \tilde{E}_2) - \tilde{d}_{l,2}, r ( \tilde{E}_2) + \tilde{d}_{r,2})$ by choosing the minimum possible value of $\tilde{d}_{r,2}$ such that $I_{r,2}( \tilde{d}_{r,2})$ is at least $t$ on the right boundary. Similarly, continuing this procedure iteratively, we compute $\tilde{E}_3.\ldots,\tilde{E}_\mathcal{M}$, resulting in the final scaled-down optimal ASK constellations, along with the left and right boundary distances.

Recall that to solve the optimization problem (\ref{eq38}), we need to find the maximum value of the error exponent $t$ for which the scaled-down optimal constellation obeys the scaled-down energy constraint. Therefore, we start with an arbitrarily large value of $t$ and keep reducing $t$ until we first encounter the value where the obtained constellation obeys the energy constraint. This is done efficiently using the bisection method shown in Algorithms~\ref{3} and \ref{4} for the one- and the two-sided ASK constellations, respectively. Finally, the scaled-down constellation is scaled-up using the expression $E_m = \mathcal{C} \tilde{E}_m$, with the final scaled-up decoding regions being $I_m = (r(E_m) - \mathcal{C} d_{l,m}, r(E_m) + \mathcal{C} d_{r,m})$ $\forall m \in \left[\mathcal{M}\right]$, with $\mathcal{M}$ and $\mathcal{C}$ defined in (\ref{eq23}) and (\ref{eq24}), respectively.

It is finally noted that the main difference between the one- and the two-sided constellation design algorithm is that the forms is initialized with energy $E_1 = 0$. In contrast, the two-sided constellation starts with a non-zero energy $E_1$. Thus, for the algorithm to work for the two-sided constellation, an extra initial energy $E_0 = 0$ is needed in the algorithmic procedure, which is later discarded. Furthermore, the scaled-up energy codebook $\mathcal{E}$ provides increased decoding region space to account for higher noise variance, improving SER performance at lower SNR while remaining under the energy constraint for traditional ASK modulation.
\subsection{Case 2: Channel Knowledge up to the Fourth Moment}
\begin{algorithm}[!t]
\caption{One-Sided ASK Constellation for Case 2}\label{alg3}
\begin{algorithmic}[1]
\State $[\mathcal{E}, \mathcal{D}] \gets \Call{BisectionOneSided}{}$
\vspace{1mm}
\hrule
\vspace{1mm}
\Function{ConstellationDesign}{$t$}
    \State Initialize $\tilde{E}_1 \gets 0$, $\tilde{d}_{l,1} \gets 0$.
    \State Set $\tilde{d}_{r,M} \gets \infty$, $\tilde{d}_{r,m} \gets \sqrt{2t\,s\left(0\right)}$.
    \For{$m = 2$ to $M$}
        \State Find smallest $\tilde{E}_m > \tilde{E}_{m-1}$ such that
        \Statex \hspace{1cm} $\dfrac{\left( \tilde{E}_m - \tilde{E}_{m-1} \right)^2}
        {2\left( \sqrt{s\left(\tilde{E}_m\right)} + \sqrt{s\left(\tilde{E}_{m-1}\right)} \right)^2} = t$.
        \If{$m \neq M$}
            \State Set $\tilde{d}_{r,m} \gets \sqrt{2t\,s\left(\tilde{E}_m\right)}$.
        \EndIf
        \State Set $\tilde{d}_{l,m} \gets \sqrt{2t\,s\left(\tilde{E}_m\right)}$.
    \EndFor
    \State Assemble $\tilde{\mathcal{E}} \gets \left\{ \tilde{E}_m \right\}_{m=1}^{M}$.
    \State Assemble $\tilde{\mathcal{D}} \gets \left\{ \left(\tilde{d}_{l,m},\, \tilde{d}_{r,m}\right) \right\}_{m=1}^{M}$.
    \State \Return $\tilde{\mathcal{E}},\, \tilde{\mathcal{D}}$
\EndFunction
\end{algorithmic}
\end{algorithm}
We now consider the case where the transceiver pair knows only the first four moments of the fading distribution. Such a relaxation in the knowledge of the channel's statistical information is relevant for practical communication systems where channel models are still under development (e.g., small-scale fading models for millimeter channels \cite{mmWave_model, mmWave_model_2, mmWave_model_3}). For such cases, the first few moments of channel fading may be easily estimated as compared to its entire statistics.

It can be seen from~(\ref{eq28}) that, for small values of $d_{l,m}$ and $d_{r,m}$, we can approximate the rate functions defined in (\ref{eq30a}) and (\ref{eq30b}) as follows:
\beqarr
I_{r,m}\left(d_{r,m}\right) \approx \tilde{I}_{r,m}\left(d_{r,m}\right) = \frac{d_{r,m}^2}{2s\left(E_m\right)}, \nn \\
I_{l,m}\left(d_{l,m}\right) \approx \tilde{I}_{l,m}\left(d_{l,m}\right) = \frac{d_{l,m}^2}{2s\left(E_m\right)},
\label{eq39}
\eeqarr
where we used the definition:
\beq
s\left(E_m\right) \triangleq \mathbb{E}\left[u_m^2\right] = \alpha_{1}E_m^2 + \alpha_{2}E_m + \alpha_{3} \, ,
\eeq
with $\alpha_1 = \mathbb{E}\left[\left(\sum\limits_{n=1}^{N} |h_{1,n}| |h_{2,n}| \right)^4\right] - 1$, $\alpha_2 = 2\tilde{\sigma}_n^2$, and $\alpha_3 = \tilde{\sigma}_n^4$. The latter follows from the fact that the noise follows a zero-mean Gaussian and is independent of the channel gains. It is crucial to note that this approximation relies solely on the first, second, and fourth moments of the fading channels.

\begin{algorithm}[!t]
\caption{Two-Sided ASK Constellation for Case 2}\label{alg4}
\begin{algorithmic}[1]
\State $[\mathcal{E}, \mathcal{D}] \gets \Call{BisectionTwoSided}{}$
\vspace{1mm}
\hrule
\vspace{1mm}
\Function{ConstellationDesign}{$t$}
    \State Initialize $\tilde{E}_0 \gets 0$, $\tilde{d}_{l,1} \gets 0$.
    \State Set $\tilde{d}_{r,M/2} \gets \infty$, $\tilde{d}_{r,m} \gets \sqrt{2t\,s\left(0\right)}$.
    \For{$m = 1$ to $M/2$}
        \State Find smallest $\tilde{E}_m > \tilde{E}_{m-1}$ such that
        \Statex \hspace{1cm} $\dfrac{\left( \tilde{E}_m - \tilde{E}_{m-1} \right)^2}
        {2\left( \sqrt{s\left(\tilde{E}_m\right)} + \sqrt{s\left(\tilde{E}_{m-1}\right)} \right)^2} = t$.
        \If{$m \neq \dfrac{M}{2}$}
            \State Set $\tilde{d}_{r,m} \gets \sqrt{2t\,s\left(\tilde{E}_m\right)}$.
        \EndIf
        \State Set $\tilde{d}_{l,m} \gets \sqrt{2t\,s\left(\tilde{E}_m\right)}$.
    \EndFor
    \State Assemble $\tilde{\mathcal{E}} \gets \left\{ \tilde{E}_m \right\}_{m=1}^{M/2}$.
    \State Assemble $\tilde{\mathcal{D}} \gets \left\{ \left(\tilde{d}_{l,m},\, \tilde{d}_{r,m}\right) \right\}_{m=1}^{M/2}$.
    \State \Return $\tilde{\mathcal{E}},\, \tilde{\mathcal{D}}$
\EndFunction
\end{algorithmic}
\end{algorithm}
To apply the approximations described in (\ref{eq39}), we need to ensure that the values of $d_{l,m}$ and $d_{r,m}$ are small enough. Thus, we use the scaled-down energy and bounds as defined in (\ref{eq37}) to re-write the optimization problem in (\ref{eq35}) as:
\begin{align}\label{eq41}
    & \underset{\{\tilde{E}_m, \tilde{d}_{l,m}, \tilde{d}_{r,m}\}_{m \in [\mathcal{M}]}}{\text{max}} 
     \quad \min_{m \in \left[\mathcal{M}\right]}\left(\tilde{I}_{l,m}\left(\tilde{d}_{l,m}\right), \tilde{I}_{r,m}\left(\tilde{d}_{r,m}\right)\right)\nonumber \\
    & \quad\quad\text{s.t.} 
    \quad  \frac{1}{\mathcal{M}}\sum_{m=1}^{\mathcal{M}}\tilde{E}_m \leq 1,  \,\, \tilde{E}_m \geq 0 \, \forall m \in \left[\mathcal{M}\right].
\end{align}
Clearly, the modified rate functions $\tilde{I}_{l,m}$ and $\tilde{I}_{r,m}$ have the same properties as the rate functions used in Section~III.A, hence, the approach to solve (\ref{eq41}) is similar to the solution for (\ref{eq38}). The only difference is that both $\tilde{I}_{l,m}$ and $\tilde{I}_{r,m}$ exhibit easily interpretable dependencies on $E_m$, $d_{l,m}$, and $d_{r,m}$. Algorithms \ref{alg3} and \ref{alg4} summarize the solution to (\ref{eq41}) for one- and two-sided ASK constellations, respectively.
\section{Numerical Results and Discussion}
This section presents numerical evaluation results for the performance metrics derived for the proposed RIS-assisted noncoherent communication. We assess the system’s error performance by considering both the traditional equispaced one- and two-sided ASK modulation schemes (indicated as `Trad.' in all performance figures), as well as our novel SER-optimal one- and two-sided ASK modulations (`Opt.' in all performance plots) for two case, one with both communication ends possessing perfect knowledge of the statistics of the channel (Case 1), and the other for knowledge of only up to the fourth moment of the channel distribution (Case 2). 
\subsection{Results for Case 1}
\begin{figure}[!t]
\centering
\subfigure[One-Sided 4-Level ASK.]
{\includegraphics[height=2.2in,width=3.4in]{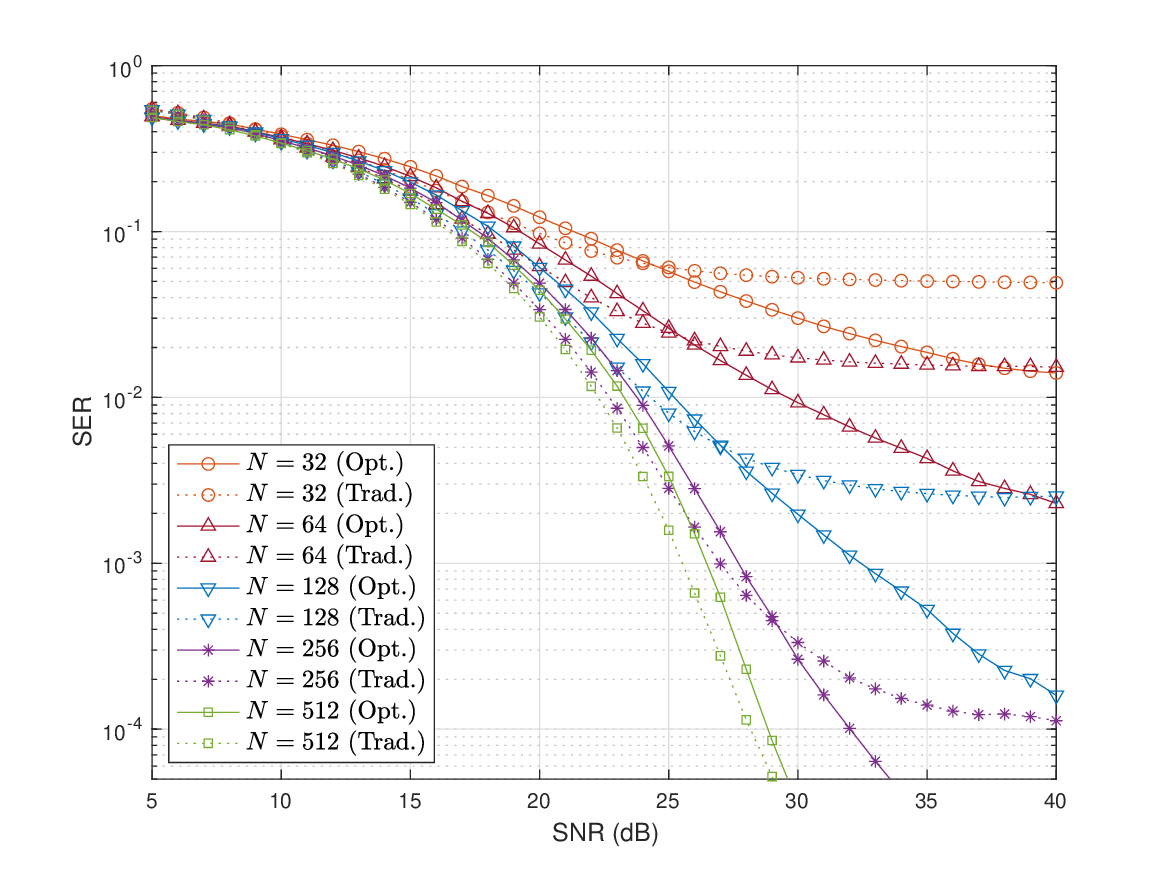}}
\subfigure[One-Sided 8-Level ASK.]
{\includegraphics[height=2.2in,width=3.4in]{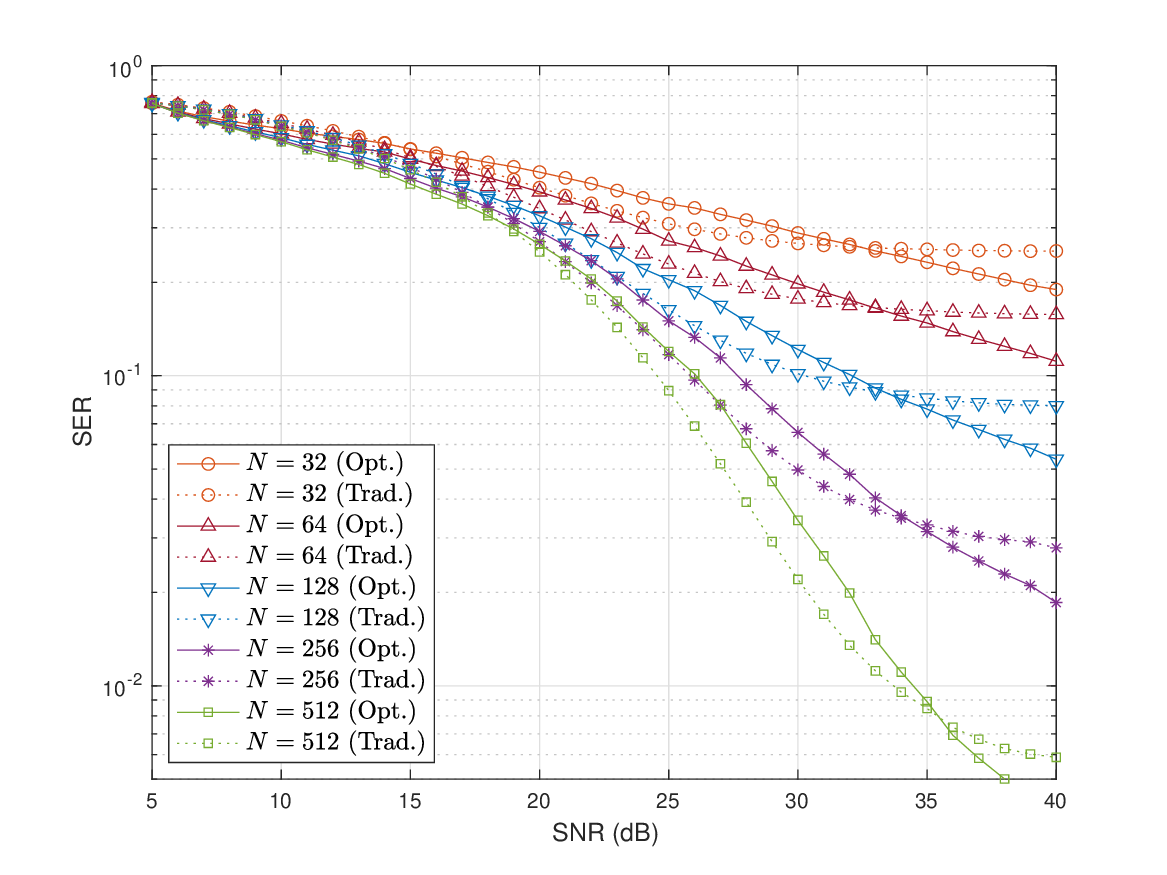}}
\caption{SER versus SNR for Case 1 with the transmitter employing one-sided (a) 4-level and (b) 8-level ASK.}
\label{f2}
\end{figure}
\begin{figure}[!t]
\centering
\includegraphics[height=2.4in,width=3.4in]{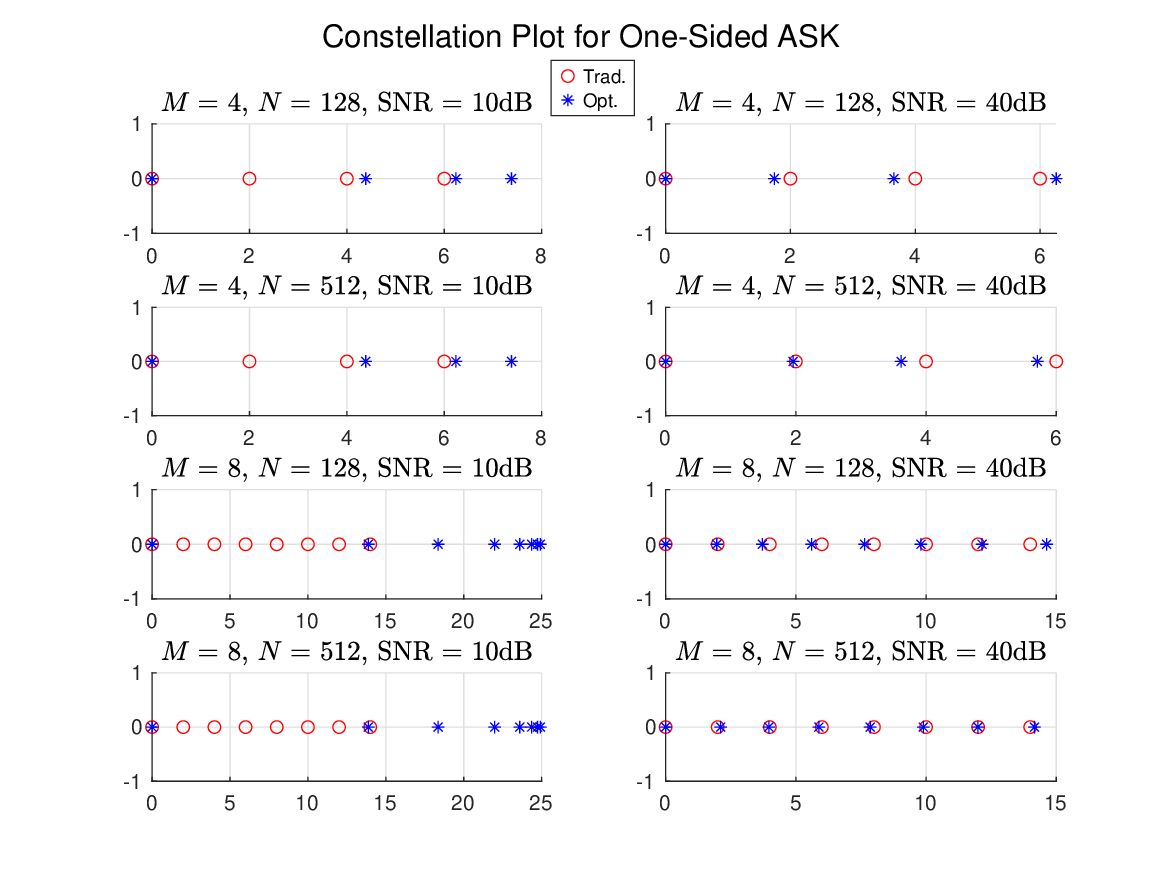}
\caption{Comparison of conventional and SER-optimal one-sided constellations for $M=\left\{ 4,8\right\}$, $N=\left\{ 128, 512 \right\}$, and SNR $=\left\{10,40 \right\}$ dB.}
\label{f3}
\end{figure}
Figure~\ref{f2} demonstrates the SER of the considered RIS-assisted SISO system for varying SNR values and number of RIS elements, considering that the transmitter employs 4- and 8-level traditional and the proposed SER-minimizing one-sided ASK constellations. It can be observed that SER improves with increasing numbers of RIS elements. Additionally, as expected, the SER degrades with the increase in the modulation order of the system. Furthermore, it is shown that, although the traditional one-sided ASK has superior performance over the optimal one-sided ASK at lower SNR values, there exists a cutoff SNR over which the performance trend is reversed and the benefit of employing our optimal ASK constellations is availed. This cutoff SNR value increases as the number of RIS elements and the modulation order used for data transmission increase. It is also depicted that the SER performance when employing the traditional one-sided ASK reaches a saturation point with increasing SNR values. On the contrary, the use of the SER-optimal one-sided ASK avoid any saturation, and thus, implying a larger diversity order compared with the conventional equivalent scheme.
\begin{figure}[!t]
\centering
\subfigure[Two-Sided 4-Level ASK.]
{\includegraphics[height=2.2in,width=3.4in]{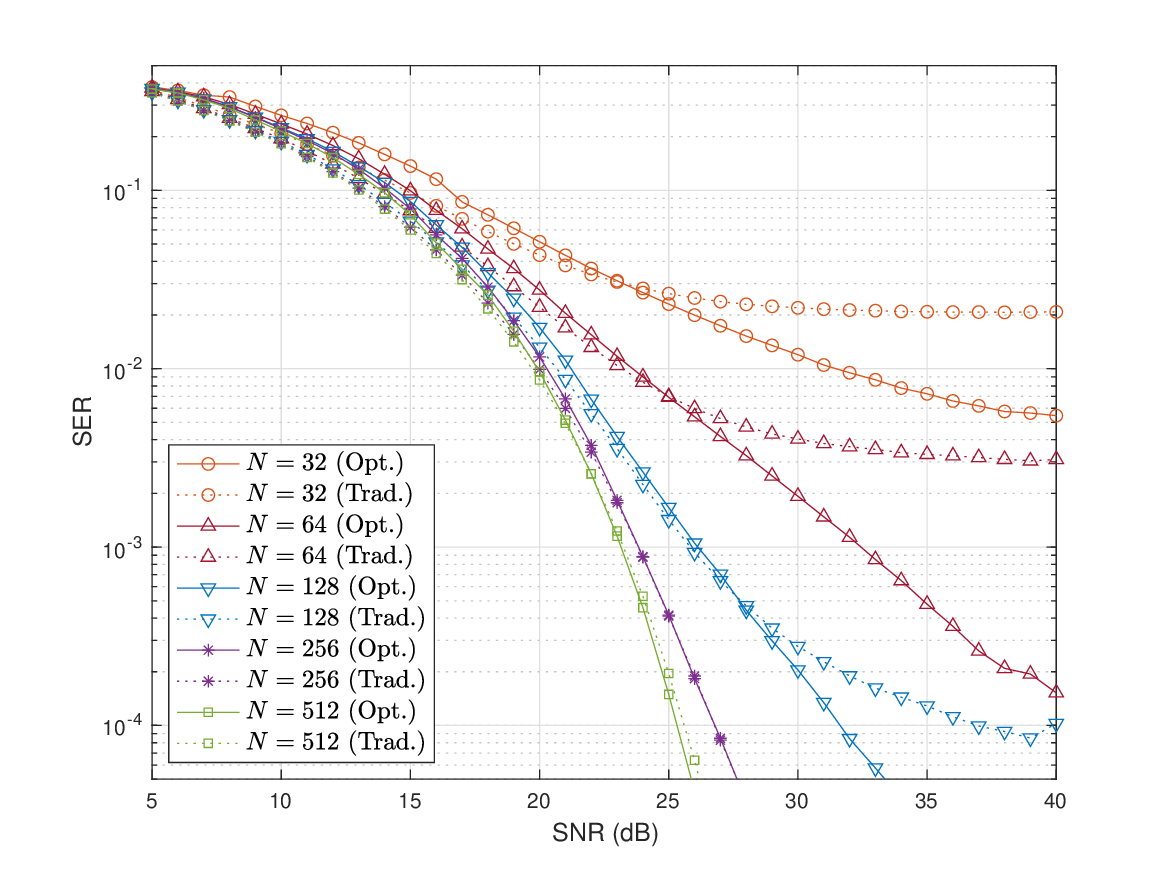}}
\subfigure[Two-Sided 8-Level ASK.]
{\includegraphics[height=2.2in,width=3.4in]{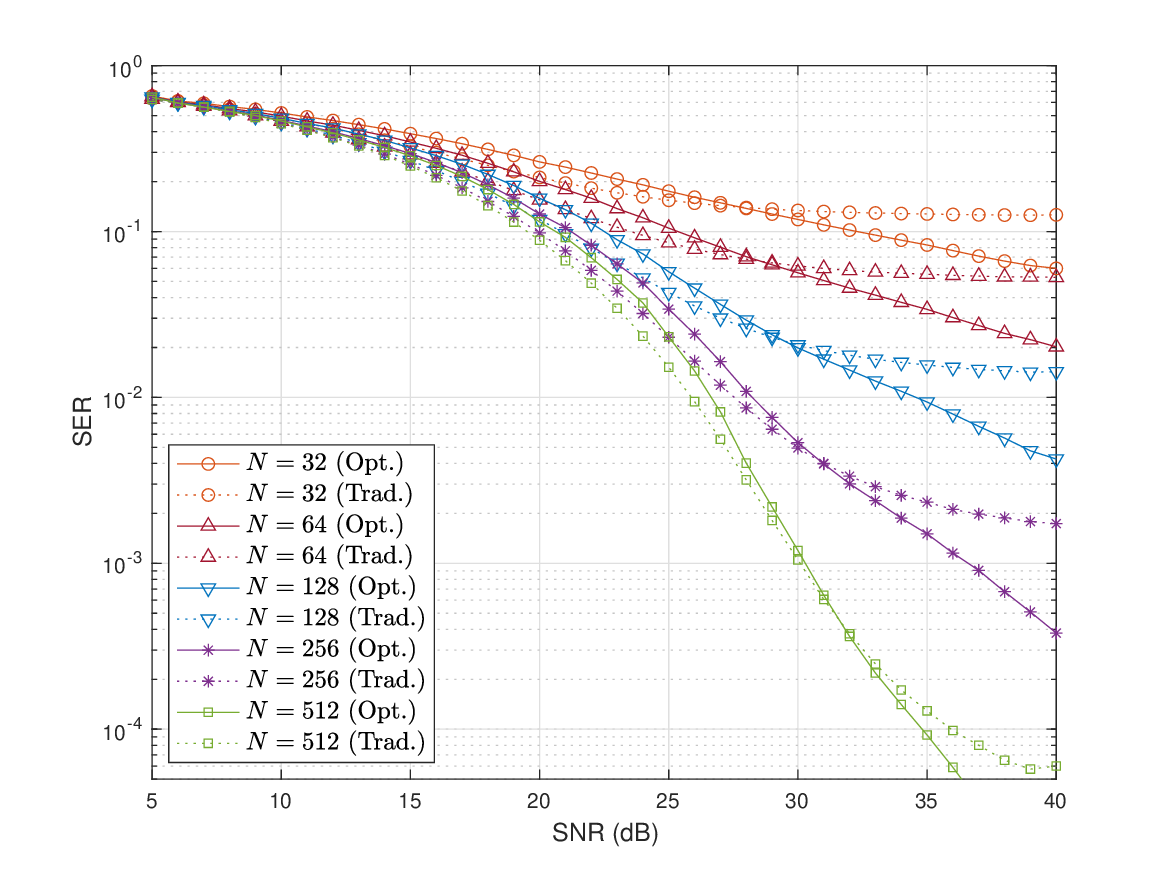}}
\caption{SER versus SNR for Case 1 with the transmitter employing two-sided (a) 4-level and (b) 8-level ASK.}
\label{f4}
\end{figure}

\begin{figure}[!t]
\centering
\includegraphics[height=2.4in,width=3.4in]{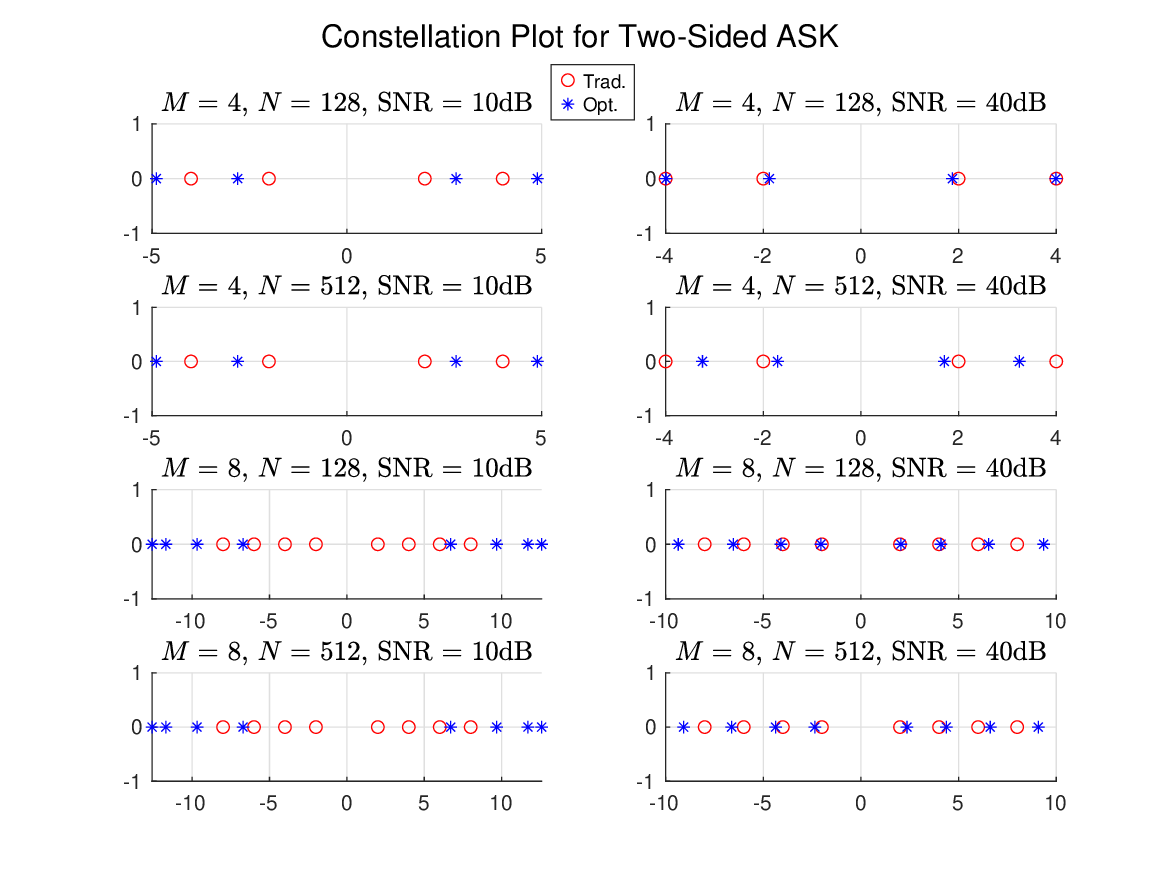}
\caption{Comparison of conventional and SER-optimal two-sided constellations for $M=\left\{ 4,8\right\}$, $N=\left\{ 128, 512 \right\}$, and SNR $=\left\{10,40 \right\}$ dB.}
\label{f5}
\end{figure}

The difference between the conventional and the proposed one-sided ASKs for varying values of $N$ is illustrated in Fig.~\ref{f3}. We have chosen two SNR values, one in the range where the performance of the traditional ASK is superior, and the other where the performance of the SER-optimal ASK is superior. It can be observed that, at low SNR values, the constellation diagrams of the schemes differ substantially. With the conventional scheme, the points are equidistant, while, with the proposed one, those with higher energies are getting concentrated at a single point. This concentration actually increases with increasing values of $M$ and $N$. On the contrary, there seems to be a smaller difference gap between the two constellations at higher SNR values. In this regime, the optimal constellations are well spread out similar to conventional ones. It can also be observed that the gap between the traditional and optimal constellations is lower for higher values of $N$.

Similar to Fig. \ref{f2}, Fig. \ref{f4} includes the SER plots versus the SNR for varying number of RIS elements with the transmitter employing 4- and 8-level conventional and the proposed SER-optimal two-sided ASK constellations. The observations are similar to that of the one-sided ASK constellations, i.e., the performance of the system improves with an increase in the value of $N$ and degrades with increasing $M$. Furthermore, the optimal constellation outperforms the traditional one after a certain threshold SNR, and this crucial SNR value increases with increasing $N$ and $M$. However, contrary to the use of one-sided ASK, the SER performance of the system employing two-sided ASK is better and the saturation effect takes place in larger SNRs for the traditional two-sided ASK modulation. It is noted that, due to the symmetry of the two-sided ASK, the optimization problem solves for $\frac{M}{2}$ levels of the constellation set, requiring less computational overhead for the execution of the optimization algorithm. Figure~\ref{f5}, similar to Fig.~\ref{f3}, compares both two-sided ASK modulations for varying values of $N$, $M$, and SNR. As shown, there is a large separation between the traditional and the optimal ASK constellations at lower SNR values as compared to higher SNRs. However, it can be seen that, at low SNR values, the location of the constellation points is less prominent than that with one-sided ASK. At high SNR values, the gap between the optimal and the traditional ASKs increases when the number of RIS elements increases. In all studied setting, it is shown that the proposed optimal two-sided ASKs move away from the origin, with this behavior depending on the values of $N$.
\subsection{Results for Case 2}
\begin{figure}[!t]
\centering
\subfigure[One-Sided 4-Level ASK.]
{\includegraphics[height=2.2in,width=3.4in]{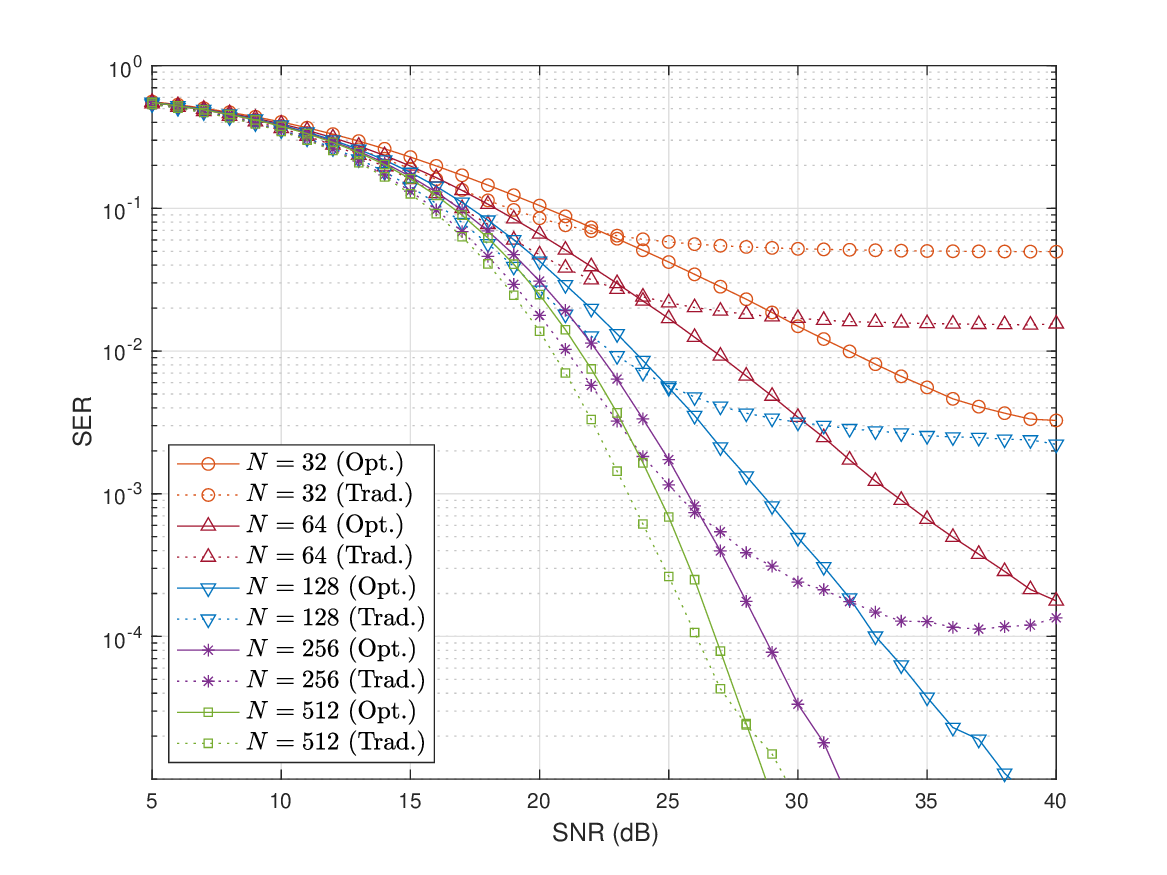}}
\subfigure[One-Sided 8-Level ASK.]
{\includegraphics[height=2.2in,width=3.4in]{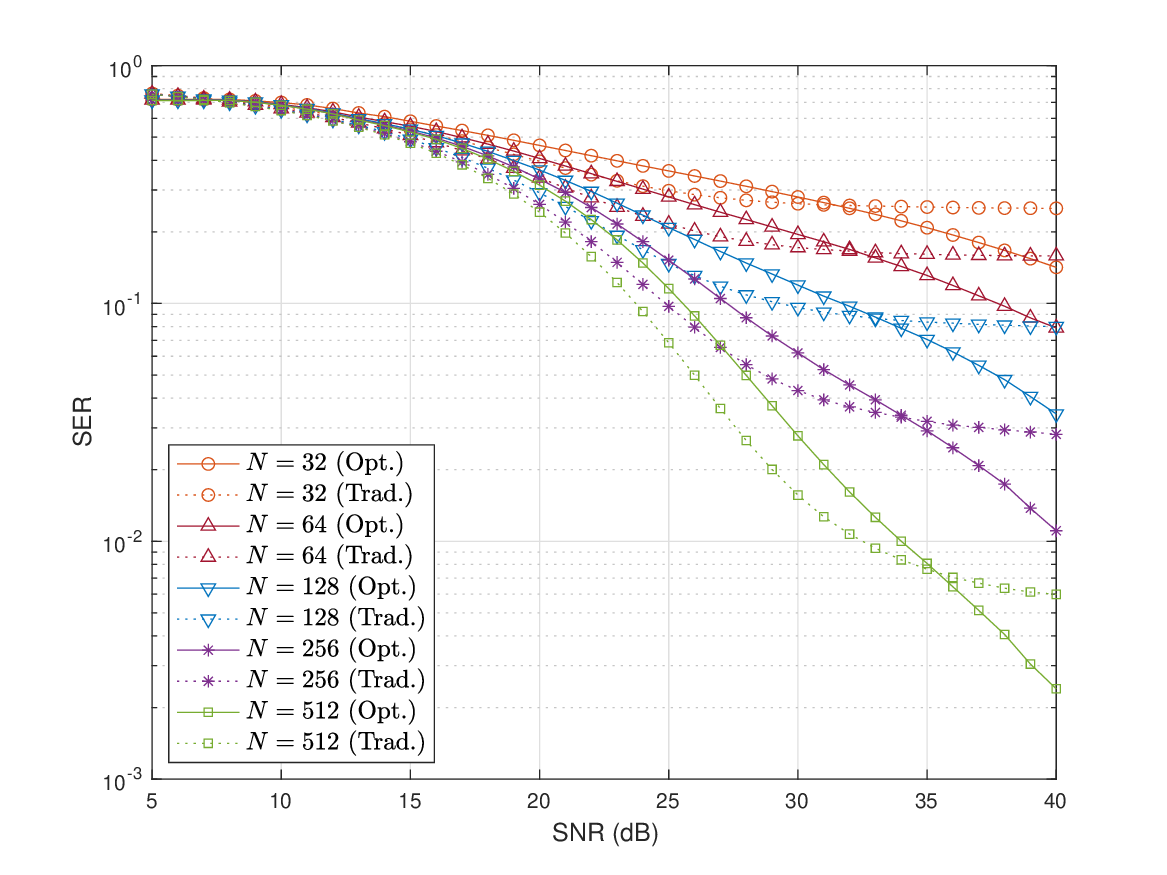}}
\caption{SER versus SNR for Case 2 with the transmitter employing one-sided (a) 4-level and (b) 8-level ASK.}
\label{f6}
\end{figure}

\begin{figure}[!t]
\centering
\subfigure[Two-Sided 4-Level ASK.]
{\includegraphics[height=2.2in,width=3.4in]{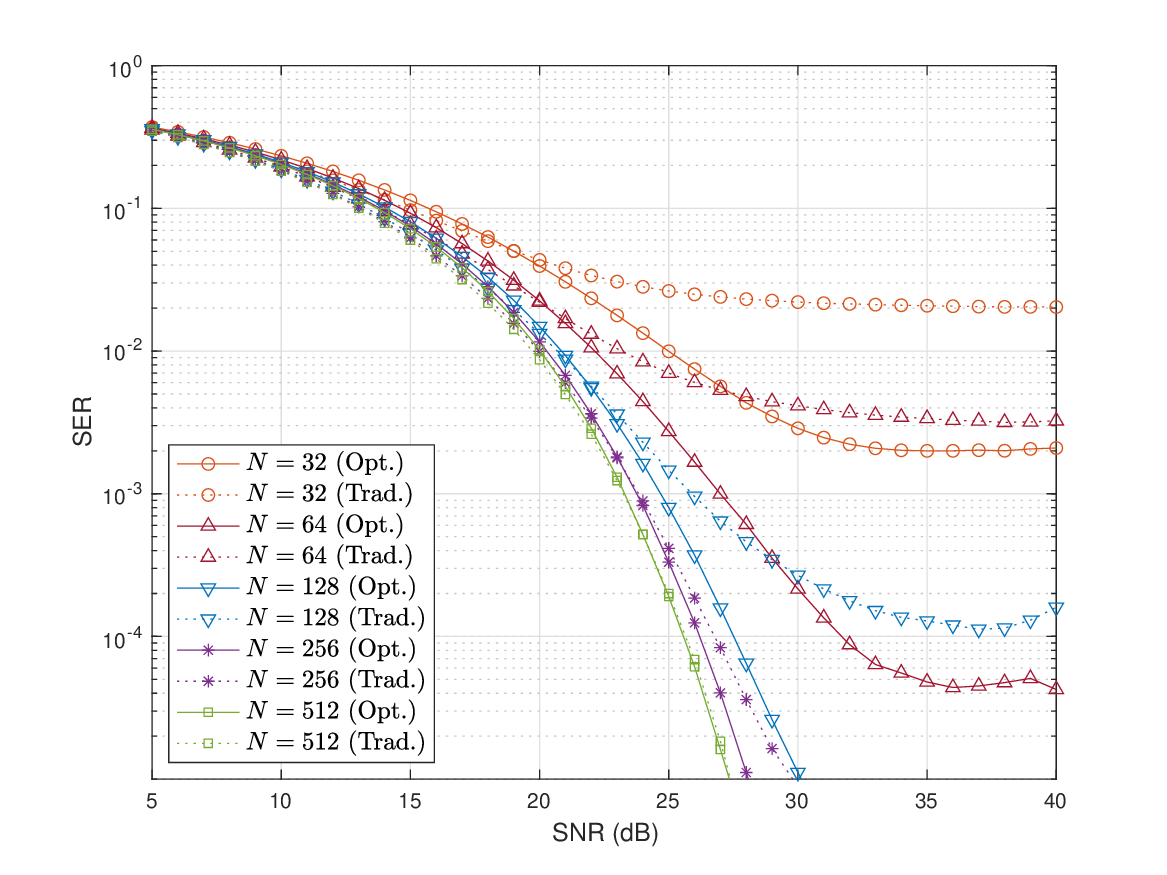}}
\subfigure[Two-Sided 8-Level ASK.]
{\includegraphics[height=2.2in,width=3.4in]{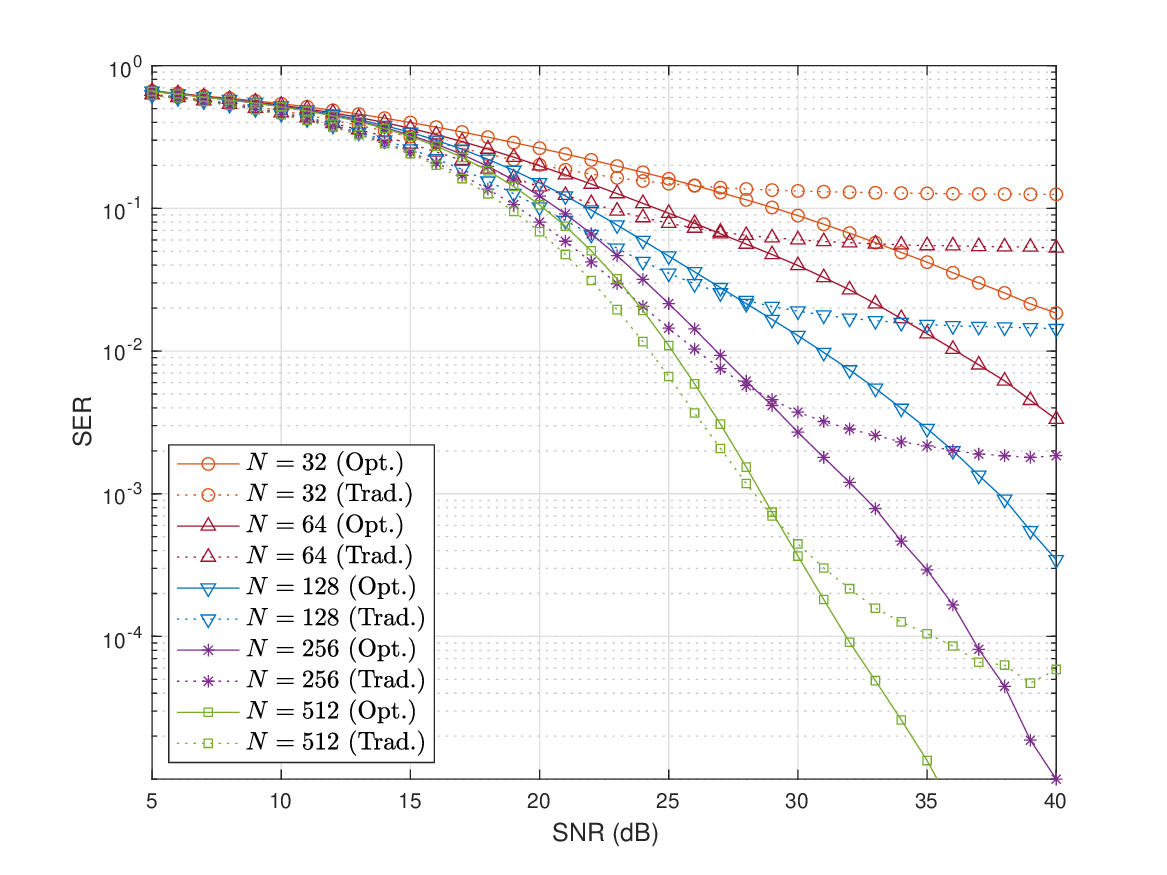}}
\caption{SER versus SNR for Case 2 with the transmitter employing two-sided (a) 4-level and (b) 8-level ASK.}
\label{f7}
\end{figure}

We now present the performance results for the second case where the transceiver pair has the knowledge of the channel statistics up to its fourth moment. Fig.~\ref{f6} depicts the SER performance of the considered RIS-aided noncoherent system versus the SNR for varying number of RIS elements when the transmitter employs one-sided 4- and 8-level ASK modulation. Similar to the results for Case 1, it can observed that there exists a threshold SNR value above which the performance of our SER-minimizing ASK outperforms that of the conventional ASK schemes. This threshold SNR is shown to increase with increasing $M$ and/or $N$. Additionally, SER increases with increasing $N$ and decreasing $M$. As expected, the SER performance for this Case 2 is worse than Case 1. However, due to the reduced computations required in this case (due to the approximations used in the computation process), the algorithm computes faster and exhibits fewer approximation errors compared to the latter scenario, and thus, it can be considered as reliable for a system requiring to operate with low computational overhead.

\begin{figure}[!t]
\centering
\includegraphics[height=2.4in,width=3.4in]{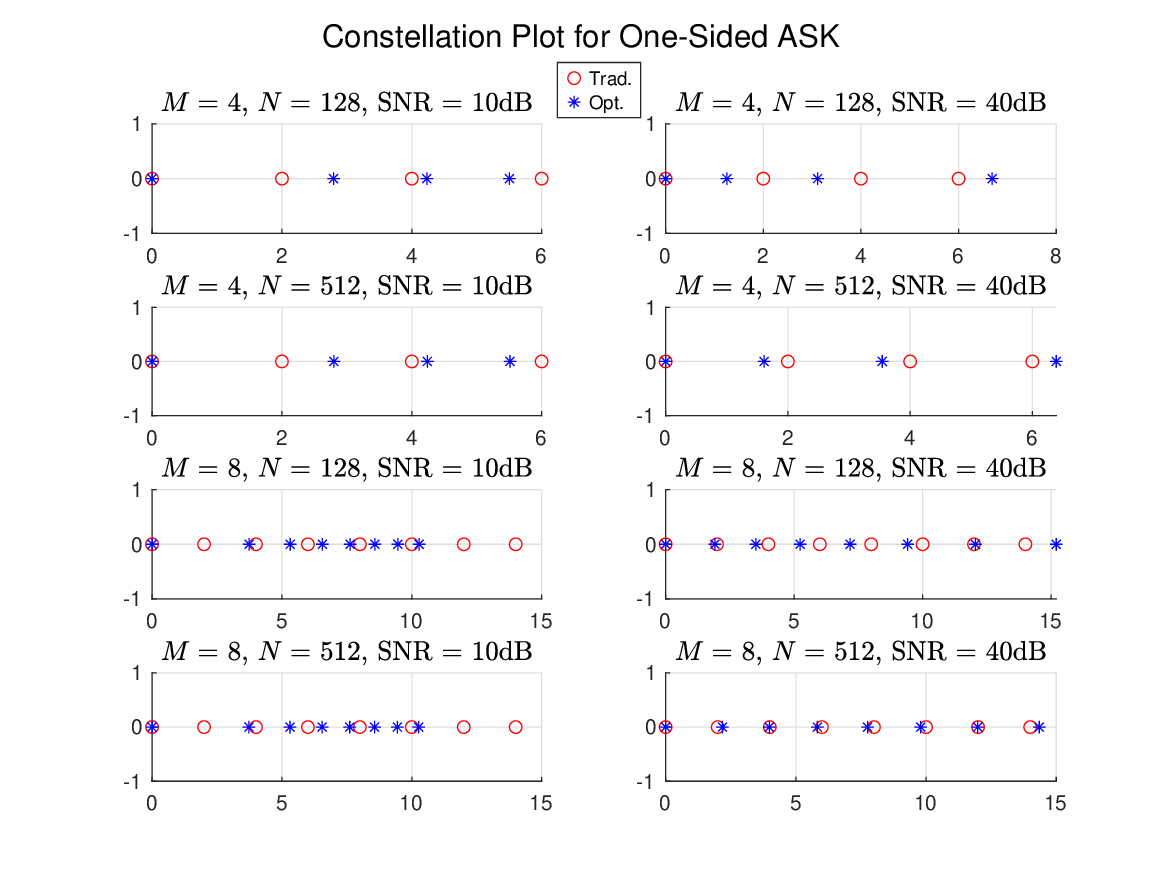}
\caption{Comparison of conventional and SER-optimal one-sided constellations for $M=\left\{ 4,8\right\}$, $N=\left\{ 128, 512 \right\}$, and SNR $=\left\{10,40 \right\}$ dB.}
\label{f8}
\end{figure}

\begin{figure}[!t]
\centering
\includegraphics[height=2.2in,width=3.4in]{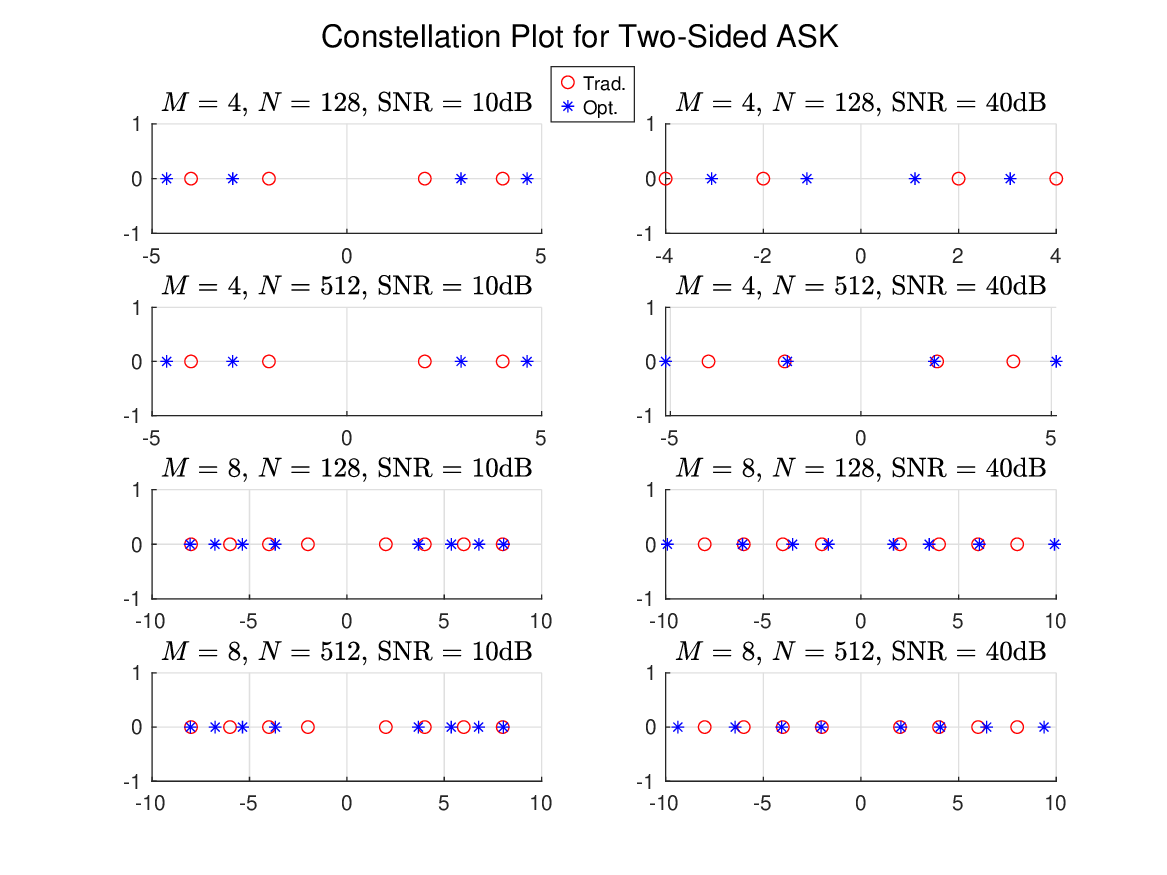}
\caption{Comparison of conventional and SER-optimal two-sided constellations for $M=\left\{ 4,8\right\}$, $N=\left\{ 128, 512 \right\}$, and SNR $=\left\{10,40 \right\}$ dB.}
\label{f9}
\end{figure}

The SER versus SNR plots for two-sided 4- and 8-level ASK modulations for varying numbers of RIS elements is shown in Fig.~\ref{f7}. The trends are similar to the previous results, with a threshold SNR value being present above which the optimal ASKs outperform the traditional ones. These thresholds increase with increasing $M$ and $N$, however, their absolute values are lower than those for one-sided ASKs. It can also be seen that traditional ASKs saturate with this effect becoming prominent at lower SNR values for the one-sided scheme as compared to the two-sided one. The conventional and proposed one- and two-sided ASKs are finally compared for this Case 2 and for various system parameters in Figs.~\ref{f8} and \ref{f9}. Similar to Case 1, the optimal ASK constellation points come closer to the traditional ASKs at higher SNR values as compared to lower SNRs. However, in contrast to Case 1, the optimal one-sided ASK constellation points become concentrated at intermediate energy values at lower SNRs. However, this trend is observed only at higher $M$ values and it becomes prominent at higher number of RIS unit elements. Similar trends are also observed for the two-sided ASK constellations. It is noted that, although the gap between the traditional and optimal ASK constellations is lower at high SNRs, there is still a difference between the two, making the proposed constellation sets the ones with the superior performance.

\section{Conclusions and Future Work}
This paper studied RIS-assisted noncoherent SISO communications when deploying one- and two-sided equiprobable ASK constellations for data modulation and energy-based symbol-by-symbol detection at the receiver. The decision regions for data decoding were obtained for noncoherent reception and an upper bound for the SER performance was presented. Considering a constraint on the average energy of the constellation for data modulation, a novel framework for the system's SER minimization was introduced leading to SER-optimal one- and two-sided ASK constellation solutions. The proposed framework considered both the case where the communication ends possess perfect knowledge of the channel statistics, as well as the case where the channel statistics are known only up to the fourth moment. The presented extensive numerical results showcased that, although traditional equispaced one- and two-sided ASK constellations constitute the best option for low SNR values, there exists a threshold SNR value above which the proposed constellations yield the best SER performance. This threshold was shown to depend on the number of RIS elements, the modulation order, and the level of channel knowledge from the system. All in all, our investigations have shown that simple receiver structures constitute promising alternatives to complex coherent designs for RIS-assisted noncoherent systems. For future work, we intend to extend our framework to MIMO settings and incorporate coding techniques to further boost the SER performance.




\FloatBarrier
\bibliographystyle{IEEEtran}
\bibliography{IEEEabrv,references}
\end{document}